\documentclass{moriond}

\def\Journal#1#2#3#4{{#1} {\bf #2}, #3 (#4)}

\def\PRD{{\em Phys. Rev.} D}

\def\be{\begin{equation}}
\def\ee{\end{equation}}
\def\bea{\begin{eqnarray}}
\def\eea{\end{eqnarray}}

\newcommand{\Photo}{}

\begin{document}
\vspace*{4cm}
\title{The two-loop bispectrum in the effective theory of large-scale structure}

\author{Tobias Baldauf$^{1}$, Mathias Garny$^{2}$, Petter Taule$^{2}$~\footnote{Speaker} and Theo Steele$^{1}$}

\address{
    $^{1}$Department of Applied Mathematics and Theoretical Physics,
    University of Cambridge, Wilberforce Road, CB3 0WA \\
    $^{2}$Physik Department T31, Technische Universit\"at M\"unchen,
    James-Franck-Stra\ss{}e 1, D-85748 Garching, Germany
}

\maketitle\abstracts{
    We study the bispectrum of large-scale structure in the EFTofLSS including
    corrections up to two-loop. We derive an analytic result for the double-hard
    limit of the two-loop correction, and show that the UV-sensitivity can be
    absorbed by the same four EFT operators that renormalize the one-loop
    bispectrum. For the single-hard region, we employ a simplified treatment,
    introducing one extra EFT parameter. We compare our results to N-body
    simulations, and show that going from one- to two-loop extends the
    wavenumber range with percent-level agreement from $k \simeq 0.08$ to
    $0.15~h/\mathrm{Mpc}$.}

\section{Introduction}

Current and near-future large-scale structure surveys are expected to return a
wealth of information that may allow for testing deviations from $\Lambda$CDM
as well as exploring alternative models. Given extended coverage of weakly
non-linear scales, a lot of attention has been devoted to constructing a robust
perturbative description.
In Standard Perturbation Theory (SPT), dark matter is described as a perfect,
pressureless fluid, modelled by the continuity and Euler equations assuming a
vanishing velocity stress tensor. The non-linear equations are solved
perturbatively in Eulerian space. However, higher order corrections do not lead
to significant improvement on weakly non-linear scales, signalling the
breakdown of perturbation theory. The issue is that the perfect fluid
description is inaccurate on the scales of interest, and non-linear evolution
on small scales produce a significant velocity dispersion that back-reacts on
the observable scales via mode-coupling. This insight has over the last decade
lead to the development of an effective field theory approach (EFTofLSS) that
systematically captures the effect on small-scale physics onto larger,
perturbative scales~\cite{bau,car}. After coarse-graining the perturbation
fields, the equations of motion contains an effective stress tensor that
encapsulates the small-scale non-perturbative effects.

Complementary to the power spectrum, higher order statistics supplement
information that can be instrumental in disentangling bias from fundamental
physical parameters as well as providing consistency checks for the EFT
parameters. In this work, we focus on the leading non-gaussian statistic, the
bispectrum, and compute for the first time the two-loop bispectrum in an EFT
framework.

\section{Effective field theory setup}

The dynamical evolution of the coarse-grained density contrast $\delta$ and
velocity divergence $\theta = \partial_i \mathbf{v}^i$ is in the EFT framework
described by the continuity equation and a modified Euler equation,
\begin{eqnarray}
    \delta'(\mathbf{k})+\theta(\mathbf{k}) & = &
        -\int \mathrm{d}^3\mathbf{q}\,\alpha(\mathbf{q},\mathbf{k}-\mathbf{q})
            \theta(\mathbf{q})\delta(\mathbf{k}-\mathbf{q})\,, \nonumber\\
    \theta'(\mathbf{k})+\mathcal{H}\theta(\mathbf{k})
        +\frac{3}{2}\Omega_\mathrm{m}\mathcal{H}^2\delta(\mathbf{k}) & = &
        -\int \mathrm{d}^3\mathbf{q}\,\beta(\mathbf{q},\mathbf{k}-\mathbf{q})
        \theta(\mathbf{q})\theta(\mathbf{k}-\mathbf{q}) - \tau_\theta(\mathbf{k})\,.
    \label{eq:eom}
\end{eqnarray}
We include only leading contributions in the gradient
expansion to the effective stress tensor $\tau_{\theta}$ and neglect stochastic
contributions (proportional to $k^4$). In the basis we work with, the
contributions to the effective stress tensors at first and second order in
powers of the fields are
\begin{eqnarray}
    \tau_{\theta}\big|_{1} & = &
        - \gamma_1 \Delta \delta_1 \,, \nonumber\\
    \tau_{\theta}\big|_{2} & = &
        - \gamma_1 \Delta \delta_2
        - e_1 \Delta \delta_1^2
        - e_2 \Delta s^2
        - e_3 \partial_i \left[ s^{ij} \partial_j \delta_1 \right] \,,
        \label{eq:tau}
\end{eqnarray}
where $\delta_1$, $\delta_2$ are the first- and second order perturbation in
the density contrast and $s^{ij}$ is the tidal tensor. We have four free EFT
parameters: $\{\gamma_1, e_1, e_2, e_3\}$. In principle one needs to write down
EFT operators up to forth order in the fields to absorb UV-divergences of the
two-loop bispectrum, however we will opt for a strategy where we do not need
to know those higher-order operators explicitly.

The perturbative solution of the equations of motion (\ref{eq:eom}) consists at
one-loop of four loop-diagram contributions and two counterterms, of which the
counterterms arises in the EFT due to the additional effective stress tensor in
the equations. We can estimate the scaling of the UV-sensitivity of the four bare
contributions by considering external wavenumbers scaling as
$|\mathbf{k}_1| \sim |\mathbf{k}_2| \sim |\mathbf{k}_3| \sim k$
while letting the loop momentum $|\mathbf{q}| $ tend to infinity. The dominant
UV-sensitivity has the form
\begin{equation}
    k^2 P_{\mathrm{lin}}^2(k) \int^{\Lambda}\,\mathrm{d}q \, P_{\mathrm{lin}}(q)
    \equiv k^2 P_{\mathrm{lin}}^2(k) \sigma_d^2(\Lambda)
    \label{eq:uv_sens}
\end{equation}
where we defined the displacement dispersion $\sigma_d^2$ and $\Lambda$ is the
cutoff. For general configurations of the external momenta $k_1$, $k_2$ and
$k_3$, the limit has a complicated functional dependence on ratios $k_i/k_j$.
Nevertheless, it can be shown that the shape dependence of the UV-limit exactly
corresponds to that of the EFT operators defined in
Eq.~(\ref{eq:tau})~\cite{bal1,ang,bal2}. Therefore, the contributions from the
UV-region can be absorbed by the corresponding counterterms.

\section{The two-loop bispectrum}

To have an EFT description for the two-loop bispectrum, we need to assess the
UV-sensitivity of the different loop contributions. At two-loop the
UV-contributions can be divided into two categories: the \emph{single-hard (h)}
region in which one of the loop momenta becomes hard, $|\mathbf{q_1}|
\to\infty$ (or equivalently $|\mathbf{q}_2| \to\infty$), and the
\emph{double-hard (hh)} region in which both loop momenta become large,
$|\mathbf{q_1}|, |\mathbf{q_2}| \to\infty$.

\subsection{Double-hard limit}

The dominant contributions in the double-hard region have the form
\begin{equation}
    k^2 P_{\mathrm{lin}}^2(k)
        \int^{\Lambda} \mathrm{d}\mathbf{q}_1 \mathrm{d}\mathbf{q}_2 \, P_{\mathrm{lin}}(q_1) P_{\mathrm{lin}}(q_2)/q_1^2
\end{equation}
in an estimated parametric scaling. Therefore, one might suspect that the
double-hard limit can be absorbed by the same counterterms as for the one-loop
above. By computing the analytical limit of the $F_6$ kernel, we show that this
is indeed the case: the shape dependence of the double-hard limit for general
configurations of external momenta corresponds precisely to the four EFT
operators defined above. We choose a renormalization scheme where we determine
the EFT parameters by fitting the one-loop bispectrum to simulations, and
remove the double-hard contribution from the two-loop correction. In other
words we use the \emph{subtracted} two-loop bispectrum defined as
\begin{equation}
    B_{\mathrm{2L}}^{\mathrm{sub}}(k_1,k_2,k_3) =
        B_{\mathrm{2L}}(k_1,k_2,k_3) - B_{\mathrm{2L}}^{hh}(k_1,k_2,k_3).
    \label{eq:2L_sub}
\end{equation}

\subsection{Single-hard limit}

To obtain a renormalized two-loop bispectrum we still need to consider the
single-hard limit. In principle, one could take one-loop diagrams with an
insertion of a one-loop EFT operator, however as this would be complicated to
compute analytically, we opt for a numerical treatment. We follow the
prescription used for the two-loop powerspectrum in Baldauf
et.~al.~(2015)~\cite{bal3}, and consider the limit
\begin{equation}
  b_{2L}^h(k_1, k_2, k_3; \Lambda)
    \equiv
    \int_{|\mathbf{q}_{2}|<\Lambda} d^3\mathbf{q}_2\,d\Omega_{q_1}
    \left[
        \lim_{q_1\to\infty}
        q_1^2 \, b_{2L}(k_1, k_2, k_3,\mathbf q_1,\mathbf q_2)
    \right]
P_{\mathrm{lin}}(q_2)\,,
    \label{eq:b_h}
\end{equation}
where $b_{2L}$ consists of the two-loop integrand except for the factor
$P_{\mathrm{lin}}(q_1)$. We can compute this integral numerically, fixing the magnitude of
$\mathbf{q}_1$ to a large value $|\mathbf{q}_1| \gg \Lambda$. The single-hard
limit contribution to the two-loop bispectrum is then
\begin{equation}
    B_{2L}^{h}(k_1, k_2, k_3; \Lambda) =
        8 \pi \, \sigma_d^2(\Lambda) b_{2L}^h(k_1, k_2, k_3; \Lambda)\,,
    \label{eq:B_h}
\end{equation}
We assume that we can renormalize the UV effectively by a shift in the value of
the displacement dispersion
$\sigma_d^2(\Lambda) \mapsto \sigma_d^2(\Lambda) + \gamma_2(\Lambda)$.
Before writing down the corresponding counterterm, we note that part of the integral in
Eq.~(\ref{eq:b_h}) is degenerate with the double-hard contribution: For
external wavenumbers much smaller than the cutoff, the integral covers a hard
region $|\mathbf{q}_2| \gg |\mathbf{k}_{1,2,3}|$ which yields another
contribution with a shape-dependence equal to that of the double-hard region.
We choose to subtract this contribution, defining $b_{2L}^{h,\mathrm{sub}}
\equiv b_{2L}^{h} - b_{2L}^{hh}$, therefore the counterterm becomes
\begin{equation}
    B_{2L}^{\mathrm{ctr}}(k_1, k_2, k_3; \Lambda) \equiv
        \gamma_2(\Lambda) \, b_{2L}^{h,\mathrm{sub}}(k_1, k_2, k_3; \Lambda)\,.
    \label{eq:2L_ctr}
\end{equation}
In total, we have five EFT parameters at two-loop:
$\{\gamma_1,e_1,e_2,e_3,\gamma_2\}$.

\section{Numerics}

To beat cosmic variance and allow for accurate calibrations of the EFT
parameters from a modest simulation volume, we use the realization based
calculation gridPT for the tree-level, one-loop, and subtracted two-loop
contributions. The single-hard limit needed for the two-loop counterterm
appears more complex to compute in gridPT, hence we compute it using Monte
Carlo integration without specializing to a realization.

In addition to the four- and five-parameter models at one- and two-loop,
respectively, we consider a simplified approach in which we assume that the
operators in Eq.~(\ref{eq:tau}) enter in the same linear combination as the
corresponding UV-contributions of SPT. Then the four parameters can be related,
leaving one free parameter. This can (naively) be extended to two-loop by
demanding $\gamma_2 = \gamma_1$.

\begin{figure}
    \centering
    \includegraphics[width=0.45\linewidth]{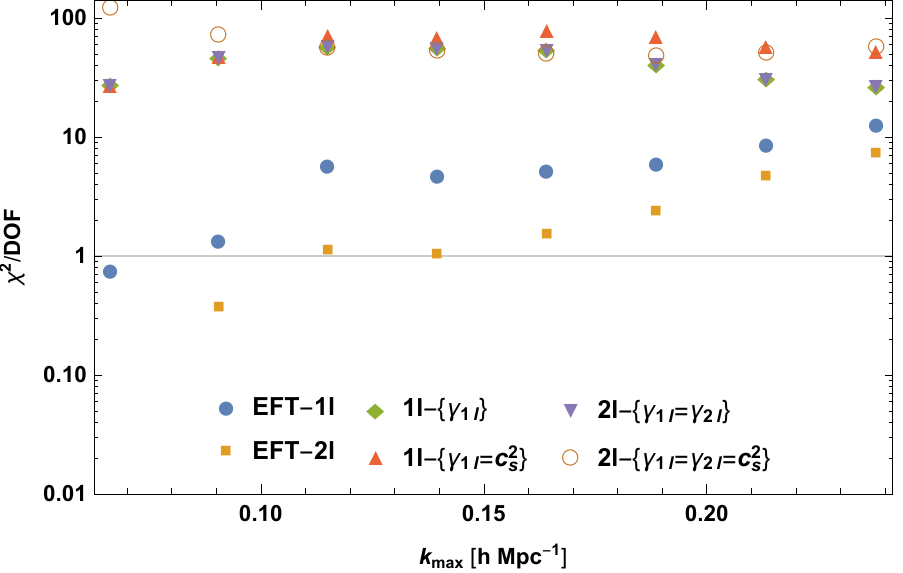}
    \includegraphics[width=0.45\linewidth]{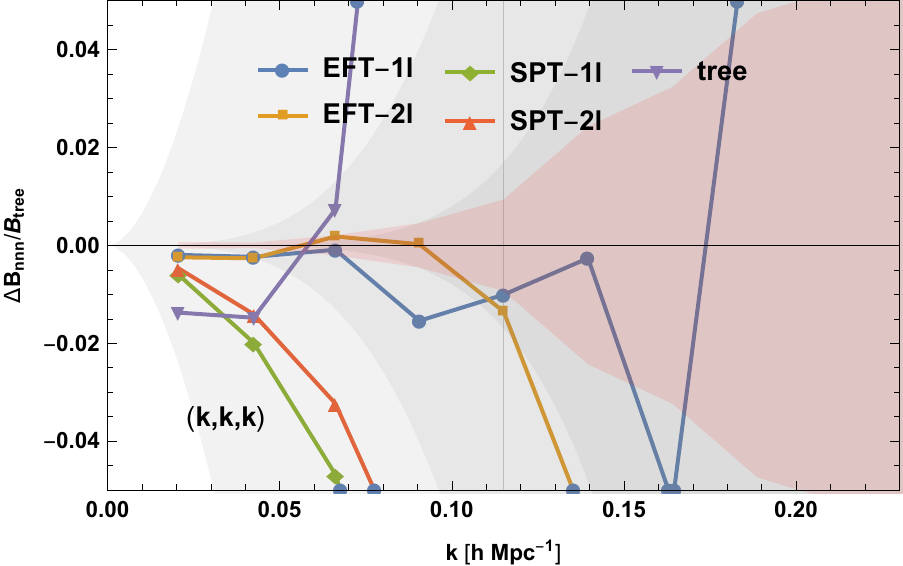}
    \caption{\emph{Left:} Reduced $\chi^2$ for the EFT one- and two-loop
    bispectra relative to N-body simulation results, for a set of triangles
    with side lengths up to $k_{\mathrm{max}}$. In addition to the four- and
    five-parameter models, we show the $\chi^2$ for simplified approaches with
    zero or one parameter (for the zero-parameter models, $\gamma_1 = c_s^2$
    where $c_s^2$ is calibrated from the power spectrum).
    \emph{Right:} The difference of the perturbative and the N-body result,
    normalized to the tree-level for a equilateral triangle configuration of
    side length $k$. The red shaded region indicates uncertainty from the
    N-body simulations, while the dark gray areas indicate expected theoretical
    uncertainty at tree-level, one-loop and two-loop, with increasing darkness,
    respectively.}
    \label{fig:chi2_pk}
\end{figure}

The reduced $\chi^2$ of the comparison with N-body simulations is shown in the
left panel of Fig.~\ref{fig:chi2_pk}. We take our full set of triangles with
$|\mathbf{k}_{1,2,3}| < k_{\mathrm{max}}$ into account, which comprises 65
(369) triangles at $k_\mathrm{max} = 0.1~(0.2)~h/\mathrm{Mpc}$. The one-loop
bispectrum remains in good fit with the N-body result for wavenumbers up to
about $k_{\mathrm{max}} \simeq 0.08~h/\mathrm{Mpc}$, while adding the two-loop
contribution extends the wavenumbers with $1\sigma$ agreement to
$k_{\mathrm{max}} \simeq 0.15~h/\mathrm{Mpc}$. It is also clear that the simple
one-parameter schemes leads to significantly larger $\chi^2$ than for the full
parametrization. In the right panel, we display the difference of the
perturbative bispectrum to the N-body result, normalized to the tree-level
bispectrum, for an equilateral configuration and with a pivot scale of
$k_{\mathrm{max}} = 0.115~h/\mathrm{Mpc}$. The EFT clearly extends the
agreement with N-body compared to SPT, and the remaining differences are
compatible with expected theoretical uncertainty. Moreover, the two-loop
correction improves the agreement compared to the one-loop result even at
relatively small scales, where the uncertainties are small.

\section{Conclusion}

In this work we compute for the first time the two-loop bispectrum of
large-scale structure in the EFTofLSS. We derive the analytic double-hard limit
of the two-loop correction, showing that this contribution can be exactly
absorbed by the four EFT operators known from the one-loop bispectrum. In
addition we adopt a simplified treatment for the single-hard region,
introducing one extra EFT parameter. We compare our results to N-body
simulations, using gridPT in order to beat cosmic variance, and find that
adding the two-loop contribution extends the range of wavenumbers with
$1\sigma$ agreement from $k\simeq 0.08$ to $0.15~h/\mathrm{Mpc}$.

\section*{Acknowledgments}

MG and PT are supported by the DFG Collaborative Research Institution Neutrinos
and Dark Matter in Astro- and Particle Physics (SFB 1258). TB is supported by
the Stephen Hawking Advanced Fellowship at the Center for Theoretical
Cosmology.

\end{document}